\pdfoutput=1      

\documentclass[aps,prl,superscriptaddress,twocolumn]{revtex4}
\usepackage{amsmath}
\usepackage{amssymb}
\usepackage[utf8]{inputenc}
\usepackage{color}
\usepackage{epsfig}

\newcommand{\un}[1]{{\,\text{#1}}}
\newcommand{\vg}{V_{\text{g}}}
\newcommand{\cg}{C_{\text{g}}}
\newcommand{\vsd}{V_{\text{sd}}}

\begin{document}

\title{Single electron tunneling through high-$Q$ single-wall carbon nanotube NEMS
resonators}

\author{A. K. H\"uttel}
\affiliation{Institute for Experimental and Applied Physics, University of Regensburg,
     93040 Regensburg, Germany}
\affiliation{Kavli Institute of Nanoscience, Delft University of Technology,
     PO Box 5046, 2600 GA Delft, The Netherlands}
\author{H. B. Meerwaldt}
\author{G. A. Steele}
\author{M. Poot}
\author{B. Witkamp}
\author{L. P. Kouwenhoven}
\author{H. S. J. van der Zant}
\affiliation{Kavli Institute of Nanoscience, Delft University of Technology,
     PO Box 5046, 2600 GA Delft, The Netherlands}

\begin{abstract}
  {\footnotesize
  \begin{center}
  \vspace*{1cm}
  \epsfig{file=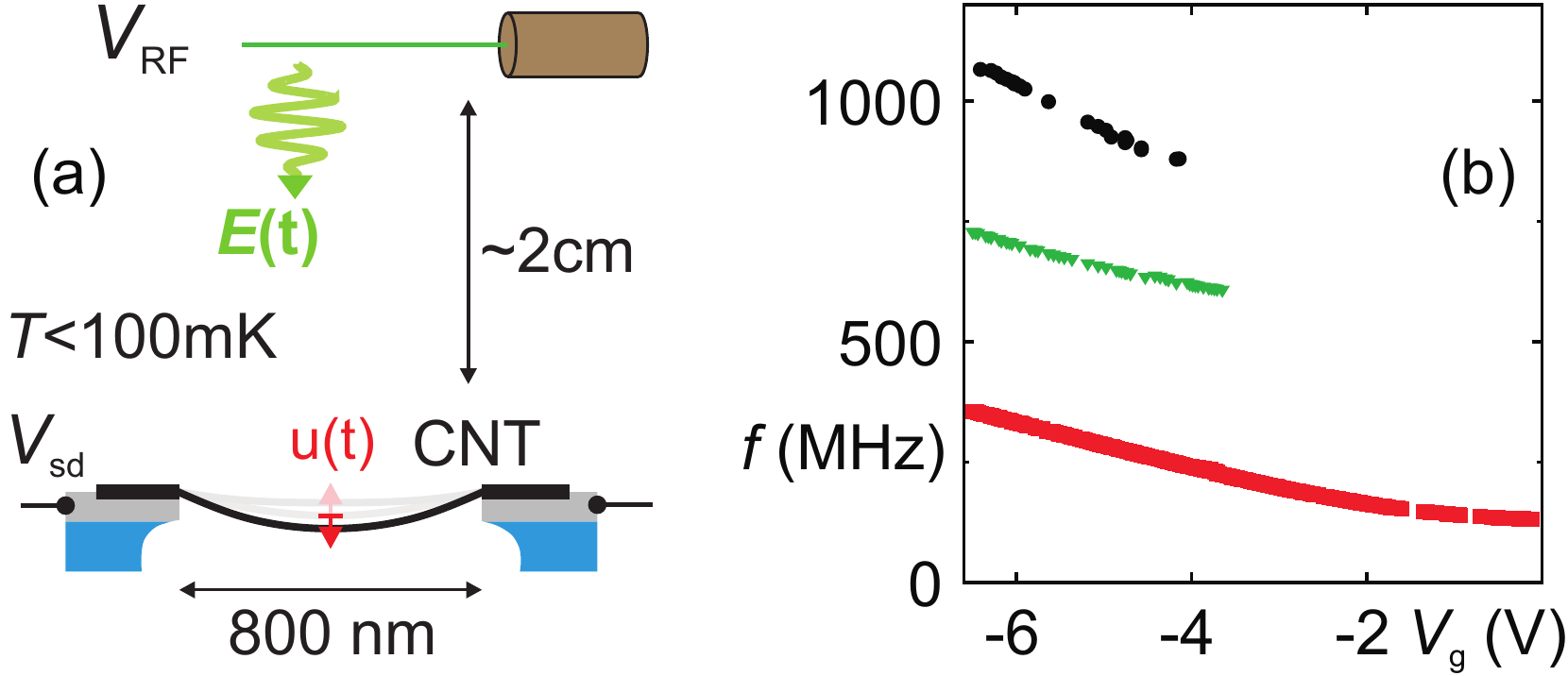,width=10cm}
  \end{center}
  (a) Sketch of the basic experimental setup, with a nanotube
  nano-electromechanical resonator being  driven contact-free via the electric field of a
  suspended radio-frequency antenna. (b) Gate voltage dependence of the observed mechanical
  resonance frequency for the fundamental bending mode of the nanotube resonator as well as
  higher modes.\\[0.5cm]}

  By first lithographically fabricating contact electrodes and then as last step
  growing carbon nanotubes with chemical vapour deposition across the ready-made chip,
  many potential contamination mechanisms for nanotube devices can be avoided. Combining
  this with pre-defined trenches on the chip, such that the nanotubes are freely suspended
  above the substrate, enables the formation of highly regular electronic systems. 

  We show that, in addition, such suspended ultra-clean nanotubes provide excellent
  high-frequency and low-dissipation mechanical resonators. The motion detection mechanism
  of our experiment is discussed, and we measure the effect of Coulomb blockade and the
  back-action of single electron tunneling on the mechanical motion. In addition
  data on the mechanical higher modes is presented.
\end{abstract}

\maketitle

\section{Introduction}

The mechanical properties of carbon nanotubes have by now inspired research for quite some
time~\cite{nature-treacy,prl-lu,prl-yu}. While applications mainly focus on bulk tensile
strength, the low mass combined with the record Young's modulus $E \simeq 1.2\un{TPa}$
also leads to extremely high resonance frequencies of doubly clamped single-wall carbon
nanotube resonators \cite{nature-sazonova:284,nl-witkamp:2904}. In the respective
nano-electromechanical devices, a single suspended carbon nanotube is clamped between
metallic leads, similar to a guitar or violin string. Motion is detected via the
electrical conductance of the sample.

As a single-wall carbon nanotube is a clean, defect-free and chemically relatively
inert macromolecule, it should act as a near-ideal mechanical beam resonator. However, for
a long time the observed
mechanical quality factors were quite low. All published experiments used basically
identical fabrication techniques -- lithographically defining contacts on top of a carbon
nanotube, followed by wet-chemical underetching -- and the same frequency downmixing
detection of mechanical motion (for details see e.g.
\cite{nature-sazonova:284,nl-witkamp:2904}). At room temperature, typical results were
$Q\lesssim 200$ \cite{nature-sazonova:284,nl-witkamp:2904,Jensen2008Atomicresolution}, 
which was increased to $Q\lesssim 2000$ in low-temperature measurements 
\cite{Lassagne2008Ultrasensitive,dilmixing}.

Whereas the detailed mechanism limiting the observable $Q$ in these measurements is still
a matter of debate, two practical points immediately come into mind: a) the fabrication
process where wet chemical treatment, resist residues, and electron beam irradiation can
lead to deterioration of the mechanical properties, and b) the measurement process where
two high-frequency signals are directly coupled into the high-impedance nanostructure,
leading to noise contributions and heating of the resonator. 

In the following sections we will address these two issues and show how optimization of 
both fabrication and measurement leads to much higher mechanical resonator quality
factors and thereby frequency resolution in clean electronic systems
\cite{highq,strongcoupling}, making a rich spectrum of interesting physics
observable.

\begin{figure*}
\epsfig{file=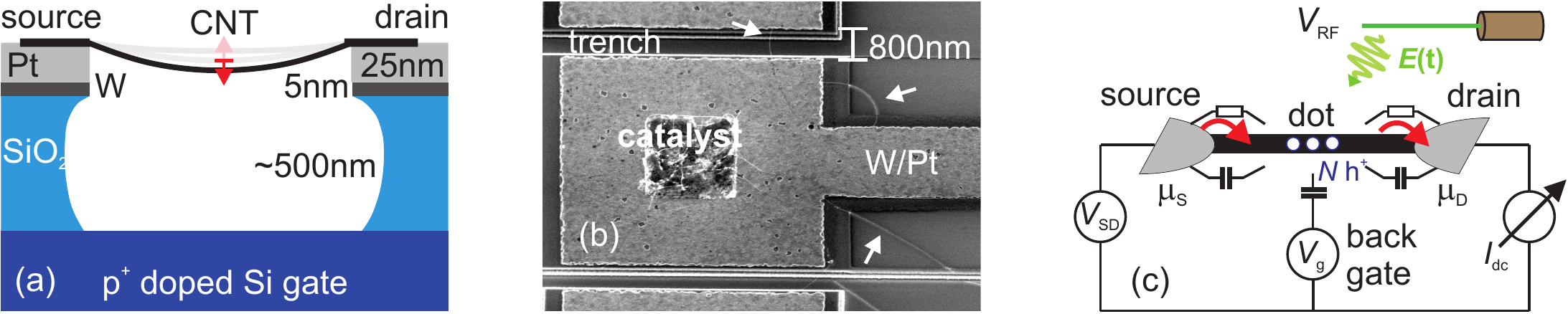,width=\textwidth}
\caption{\label{fig-sample}
  (a) Simplified side-view sketch of the chip geometry (for more details, see
  \cite{nnano-steele:363}). (b) Scanning electron microscope top-view image of a chip 
  from the same fabrication run as the measured one. The white arrows mark several
  nanotubes grown across the chip. (c) Schematic of the measurement wiring,
  integrating a contact free radio frequency antenna into a typical dc
  Coulomb blockade setup.}
\end{figure*}
\section{Device fabrication}

Figure~\ref{fig-sample}(a) displays a simplified sketch of the chip geometry used in our
experiments; a scanning electron microscope picture of a device from the same fabrication
process as the measured one is shown in Fig.~\ref{fig-sample}(b).
The full details of the device layout and the fabrication process can be found in
\cite{nnano-steele:363}. The important modification of the fabrication process
compared to previous measurements is that all lithographic processes for definition of
on-chip structures are completed before carbon nanotube growth. This includes the
definition of tungsten/platinum electrodes as well as the etching of the surface layers to
generate a trench over which the carbon nanotubes will be suspended. After localized
growth catalyst deposition, the devices are then finshed in a chemical vapour deposition
step, see \cite{nature-kong:878}. In this step, nanotubes grow across the electrodes and
the pre-defined trench between them, which on this particular sample spans $800\un{nm}$.
Since with this technique, as pioneered in \cite{nmat-cao:745}, no lithography or wet
chemistry takes place after nanotube growth, the macromolecules remain chemically and
physically undamaged and free from resist residues or other contamination.

In addition to providing better mechanical properties, the electronic properties 
of the nanotube structures improve as well. This directly follows from the 
absence of fabrication-induced defects and charges trapped at surface contaminations. 
While Ref.~\cite{highq} is the first observation of the mechanical properties of
such devices, several recent studies have already used them for fundamental research
on trapped electronic systems~\cite{nmat-cao:745,nphys-deshpande:314,nature-kuemmeth:448}.

\section{Dc current detection of mechanical resonance}

\subsection{Measurement setup}

After completion of the fabrication steps, the device was placed in a dilution
refrigerator, contacted electrically via low-pass filtered dc cabling only. 
The measurement is described schematically by the sketch of
Fig.~\ref{fig-sample}(c). At the dilution refrigerator base temperature of
$T_\text{mc}\simeq 20\un{mK}$, the
electronic system within the suspended carbon nanotube acts as a highly regular quantum
dot, with the four-fold orbital level degeneracy typical for a clean single-wall carbon
nanotube \cite{prl-oreg:365}. 
\begin{figure}[ht]%
\includegraphics*[width=\linewidth]{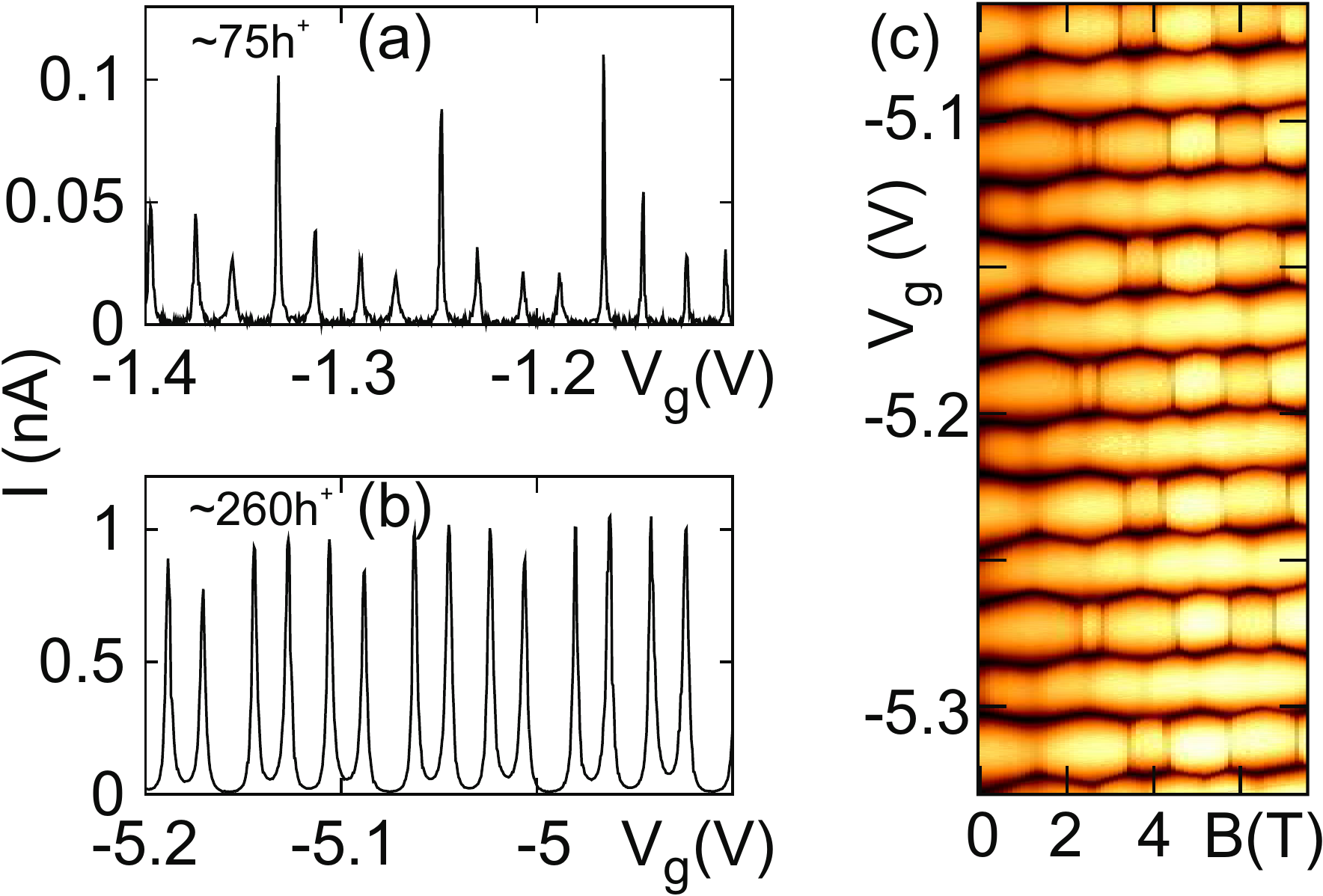}
\caption{%
  (a), (b) Exemplary measurements of
  $I_{\text{dc}}(V_g)$ at $V_{\text{sd}}=0.05\un{mV}$ for gate voltage regions with
  (a) weak and (b) strong coupling of the quantum dot to its leads. Hole numbers are
  estimated by counting the Coulomb blockade oscillations from the band gap.
  (c) Current $I_{\text{dc}}$ as function of gate voltage $\vg$ and a magnetic field $B$ 
  with a finite component parallel to the nanotube axis. The bias voltage is
  $\vsd=0.03\un{mV}$. The high regularity of the transport spectrum can be
  seen to extend to ground state orbital or spin transitions in a wide field range
  \cite{prl-jarillo:156802,nature-jarillo:484}.}
  \label{fig-setprops}
\end{figure}
This is demonstrated in Fig.~\ref{fig-setprops}(a,b), displaying $I_{\text{dc}}(V_g)$ in
the linear conductance regime. The device can be
tuned from very weak (a) to strong coupling (b) between the quantum dot and its
contacts by application of a back gate voltage alone, making both strong Coulomb blockade
and
the Kondo regime \cite{nature-goldhaber:156,prl-goldhaber:5225} accessible. As shown in
Fig.~\ref{fig-setprops}(c), the regularity of the transport spectrum extends to level
spectroscopy in a magnetic field, where multiple indications of orbital and
singlet-triplet Kondo effects at ground state transitions, i.e. level degeneracy, can be
observed \cite{prl-jarillo:156802,nature-jarillo:484}.

As already mentioned, previously the mechanical resonance of suspended nanotubes
has been detected by means of the so-called frequency downmixing technique
\cite{nature-sazonova:284,nl-witkamp:2904,Lassagne2008Ultrasensitive,dilmixing}.
This requires the application of two high frequency signals directly to the nanotube
device. The minimal temperature is thus limited by the thermal coupling of the
high frequency leads to the cold finger of the cryostat, as well as by heating effects
induced by the direct introduction of high-frequency signals. Nevertheless, an improvement
of the quality factor up to $Q\simeq 2000$ at millikelvin temperatures has been
observed~\cite{dilmixing}. In contrast, in the measurements described here, only one
high-frequency driving signal is applied contact-free -- i.e., as radiation from an
antenna. The antenna, consisting of the open end of a coaxial cable with its shield
removed, is suspended freely in the inner vacuum chamber of the dilution refrigerator at a
distance of approximately $2\un{cm}$ from the sample surface.

\subsection{Observation of the mechanical bending mode}

\begin{figure}[ht]%
\includegraphics*[width=\linewidth]{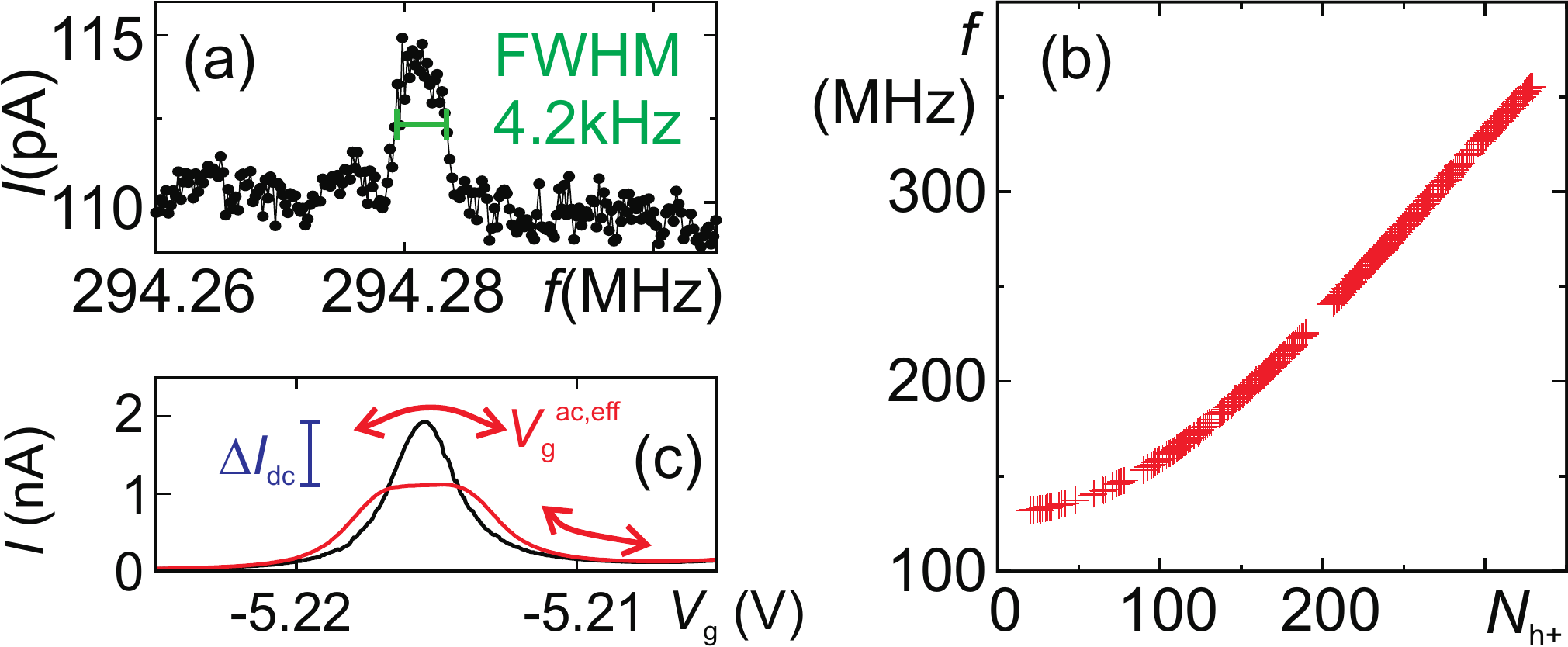}
\caption{%
  (a) Measured single electron tunneling current $I_{\text{dc}}(f)$ as function 
  of the antenna driving signal frequency, for fixed gate voltage $\vg=-5.17\un{V}$, 
  bias voltage $\vsd=0.092\un{mV}$, and nominal driving power $-60\un{dBm}$. (b)
  Dependence of the observed resonance frequency on the charge
  influenced on the nanotube, over a large gate voltage and thereby charge range. The
  number of holes on the  nanotube $N_{\text{h}^+}$ is obtained by counting the Coulomb
  blockade oscillations from the band gap. (c) Sketch of the rectification mechanism
  leading to the observed dc signal at resonance (see text).}
  \label{fig-resonance}
\end{figure}
When measuring the dc single electron tunneling current through the carbon nanotube
quantum dot while sweeping the frequency of the applied driving signal, a resonant feature
emerges even at low driving power. Figure~\ref{fig-resonance}(a) displays a typical 
example for this frequency response at $V_g=-5.17\un{V}$, $\vsd=-92\un{$\mu$V}$ and a
nominal driving power of $-60\un{dBm}$. The sharpness of the response in frequency, with a
full width at half maximum in this case here of $\Delta f=4.2\un{kHz}$ and values observed
as small as $f_0/150000$ at center frequency $f_0$, indicates a high quality factor of
the underlying resonator system. Fig.~\ref{fig-resonance}(b) plots the
gate voltage dependence of this resonance frequency over the entire measured range of
$\vg$, scaled as function of the number of holes charged on the nanotube quantum dot
$N_{\text{h}^+}$. The observed shift indicates that indeed a mechanical resonance is
observed, since in this case the resonance frequency is tuned by the electrostatically
induced 
tension \cite{nature-sazonova:284}. A comparison with a model for the bending mode of a
bulk beam resonator \cite{nl-witkamp:2904,pssb-poot:4252,highq} confirms the mechanical
origin of the current peak (see \cite{highq}). At low gate voltage the model agrees with
the data very well for $f^\ast = 132.0 \un{MHz}$ and $\vg^\ast=-2.26\un{V}$, where
$f^\ast$ is the resonance frequency in absence of tension and $\vg^\ast$ marks the
crossover between weak and strong bending regime \cite{highq}. At high negative gate
voltages $\vg\lesssim -4\un{V}$, model and measurement diverge in the region of high
mechanical tension. The origin of this is still to be determined, however it may be
related to more complex (i.e. partly longitudinal) motion of the nanotube resonator
or to limitations of the model approximations for the bending rigidity.

\subsection{Motion detection mechanism}

While the phenomenological behaviour of the observed resonance clearly indicates the
nanotube bending vibration mode as its origin, it has not been adressed so far how the
mechanical resonance actually becomes visible in dc current. Fig. \ref{fig-resonance}(c)
illustrates a straightforward mechanism. The black line depicts a measured Coulomb
blockade oscillation of current $I_{\text{dc}}(\vg)$ in absence of any radio-frequency
signal. Conventionally in such a plot the gate voltage $\vg$ is chosen as x-axis
parameter, however the parameter actually entering the equations is the product of the
voltage and the gate capacitance, $\vg \cg$. 

At resonant driving of the nanotube, the amplitude of the driving-induced oscillatory
motion of the nanotube resonantor is strongly enhanced. This motion of the quantum dot
with respect to the chip substrate modifies $\cg$, and thereby also the potential on the
dot induced by the back gate electrode. Thereby, the mechanical motion of the nanotube
acts equivalently to an ac gate voltage superimposed on the statically applied dc gate
voltage $\vg$. 

The mechanical resonance frequencies in the MHz range are far above the bandwidth of the
dc amplifiers used at the current detection side. The measurement thus averages over the
oscillation period, and thereby over the potential dependence of the source-drain
current. Since this dependence is -- due to the Coulomb blockade oscillations -- strongly
nonlinear, the peak shapes are smeared out. This is indicated by the red line in Fig.
\ref{fig-resonance}(c), obtained by numerical averaging from the measured data.
The averaging at mechanical resonance this way leads to an effective dc current
contribution $\Delta I_{\text{dc}}(\vg)$, indicated in the plot for the case of the
current maximum. Recently, a very similar mechanism has also been observed in a suspended 
resonator containing an aluminum single-electron transistor \cite{pashkin}.

Assuming an arbitrary non-linear function $I(u,\vg)$ and a harmonic oscillation of the
mechanical system with deflection $u(t)=u_0 \sin(2\pi f t)$,  we obtain for the current
time-averaged over the oscillation period $\tau=1/f$
\begin{multline} \label{current}
\overline{I}(u_0, \vg)  = \frac{1}{\tau} \int_0^\tau I(u_0 \sin(2\pi t/\tau),
\vg)\,\text{d}t
\\
= I(0, \vg) + \overline{\Delta I} (u_0, \vg)
\end{multline}
The change in current $\overline{\Delta I} (u_0, \vg)$ due to the mechanical motion with
amplitude
$u_0$ can be estimated by approximating the function $I(0,\vg)$ as a Taylor series up
to second order in $\vg$, and by linearizing the geometrical capacitance $\cg(u)$ around
the equilibrium position of the nanotube $u=0$. We then obtain as function of the motion
amplitude $u_0$
\begin{equation} \label{currentchange}
\overline{\Delta I} (u_0, \vg)= \frac{u_0^2}{4}\,
\left(\frac{\vg}{\cg} \frac{\text{d}\cg}{\text{d}u} \right)^2
\frac{\partial^2 I}{\partial \vg ^2} + \mathcal{O}\left( u_0^{4}
\right).
\end{equation}

Note that we are assuming here and in later argumentation that the single electron
tunneling current is a well-defined function of the nanotube position at each point in
time of its motion $I(u(t))$. This means that electronic tunneling into and out of the
nanotube quantum dot takes place at timescales much faster than the mechanical
oscillation. The assumption agrees well with our experiment; for weaker and
weaker coupling between quantum dot and leads the mechanical signal loses amplitude and
at some point vanishes. At the last Coulomb blockade oscillations
where it can still be observed ($V_{\text{g}}=-0.08\un{V}$), the current in large-bias
single electron tunneling ($V_{\text{sd}}=1.5\un{mV}$) reaches values of $I\simeq
50\un{pA}$. This current corrsponds to a lower bound for the tunnel frequency of
$f_t\simeq 300\un{MHz}$, compared to a mechanical resonance frequency in this gate region
of $f\simeq 132\un{MHz}$.

\subsection{Mechanical quality factor}

The quality factor of a resonator is defined as the ratio between the resonator's
center frequency and its bandwidth, i.e. the full width half maximum {\em of the stored
energy} of the steady state oscillations as function of frequency $f$. In the case of a
mechanical harmonic oscillator, the energy scales with the square of the oscillation
amplitude; using eq. \ref{currentchange} we obtain
\begin{equation}
 E_{\text{mech}}(f) \propto u_0^2(f) \propto \overline{\Delta I}(u_0(f))
\end{equation}
This way, the resonator quality factor equals the ratio of center frequency $f_0$ and
full width half maximum {\em of the measured current peak}, enabling straightforward
determination. As already mentioned, values of up to $Q\simeq 150000$ have been observed;
for the details we would like to refer to \cite{highq}.

\section{Mechanical motion and single electron tunneling}

\begin{figure*}[ht]%
\includegraphics*[width=\linewidth]{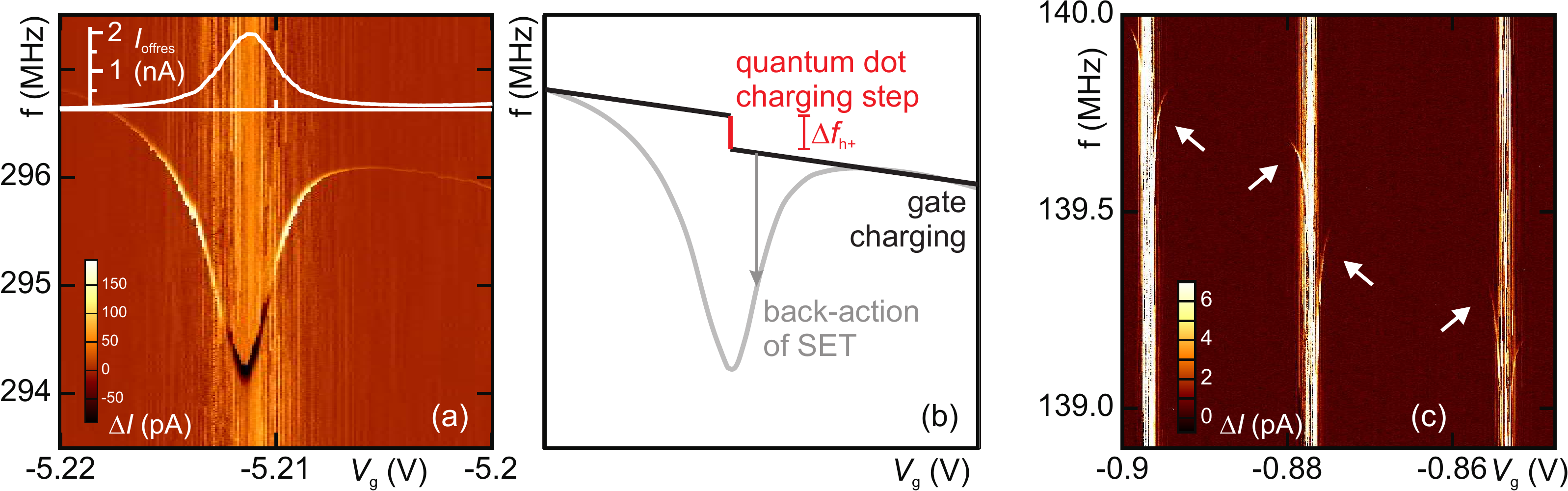}
\caption{%
  (a) High resolution measurement of the mechanical resonance signal across one Coulomb
  blockade oscillation: dc current $I_{\text{dc}}(\vg,f)$ as function of gate voltage
  $\vg$ and driving frequency $f$, for a comparatively strong radio-frequency signal of
  $-43\un{dBm}$ and $\vsd=0.1\un{mV}$. For better contrast, the average current value of
  each column has been  subtracted. Line plot: total dc current $I_{\text{dc}}(\vg)$ for
  off-resonant driving, re-using the x-axis of the main plot. 
  (b) Sketch of the structure observed in (a), with the influence of continuous back gate 
  charge, discrete quantum dot charge and single electron tunneling indicated.
  (c) $I_{\text{dc}}(\vg,f)$ as function of gate voltage $\vg$ and driving frequency $f$
  as in (a), but across three Coulomb blockade oscillations in a gate voltage region with
  much weaker dot--leads coupling. Driving power setting $-33\un{dBm}$, bias voltage
  $\vsd=0.1\un{mV}$. The arrows indicate the mechanical resonance signal emerging at the
  edges of each single electron tunneling region.
  }
  \label{fig-back}
\end{figure*}
So far, the detailed nature of Coulomb blockade and single electron tunneling has not
been taken into account; for the motion detection mechanism any nonlinear
gate voltage dependence of the measured dc current is sufficient. In addition, the plot of
Fig.~\ref{fig-resonance}(c) displays the gate voltage dependence of the observed
mechanical resonance frequency only with a limited resolution. Zooming in, 
Fig.~\ref{fig-back}(a) shows the mechanical resonance signal across
a single Coulomb blockade oscillation, providing a resolution in gate voltage
higher by a factor 250 than Fig.~\ref{fig-resonance}(c). The plotted signal is the dc
current $I_{\text{dc}}(\vg,f)$ as function of gate voltage $\vg$ and driving frequency
$f$. For each data column ($\vg=\text{const.}$) the average current value has been 
subtracted to increase the contrast of the sharp resonance peak in
$I_{\text{dc}}(f)$. The bias voltage is set to $\vsd=0.1\un{mV}$, such that the quantum
dot passes through regions of both single electron tunneling and Coulomb blockade. This is
illustrated by the white line plot insert, which displays the dc current
$I_{\text{dc}}(\vg)$ measured for off-resonant driving frequency, using the same x-axis
as the main plot. 

Across the single electron tunneling region a clear minimum of the mechanical resonance
frequency $f_0$ can be observed. In Coulomb blockade, $f_0$ approaches a linear dependence
on gate voltage, however the tangential lines are shifted for different charge on the
quantum dot. Fig.~\ref{fig-back}(b) sketches the detailed behaviour as it repeats itself
for each Coulomb blockade oscillation over a large range of the quantum dot charge. As we
will see, different details of Figs.~\ref{fig-back}(a,b) correspond to different aspects
of the coupled electromechanical system.

\subsection{Integer charge on nanotube}

In room temperature measurements a smooth and monotonous dependence of the mechanical
resonance frequency on the gate voltage is both expected and observed
\cite{nature-sazonova:284,nl-witkamp:2904}. In contrast, at low temperatures
and in the Coulomb blockade regime the charge located on the nanotube is discrete, and
can only change by integer multiples of the elementary charge. In
\cite{Sapmaz2003Carbon} it was already predicted that this should lead to a steplike
behaviour of the mechanical resonance frequency. The step size $\Delta
f_{\text{h}^+}$, as indicated in the sketch of Fig.~\ref{fig-back}(b), corresponds to the
increase in tension due to addition of a single elementary charge on the nanotube, whereas 
the finite slope between the steps is caused by the continuously varying charge on the back 
gate electrode. 

In a simple picture the electrostatic force on the nanotube per additional elementary 
charge should grow linearly with the electric field between nanotube and gate and thereby
the charge on the back gate electrode. Indeed, a clear gate voltage dependence of the
frequency step size has been observed, with values varying from $\Delta f \simeq
0.1\un{MHz}$ at $\vg \simeq -0.9\un{V}$ ($N_{\text{h}^+}\simeq 55$) to $\Delta f \simeq
0.5\un{MHz}$ at $\vg \simeq -5.2\un{V}$ ($N_{\text{h}^+}\simeq 265$). A detailed analysis
of the functional dependence would however require additional data.

\subsection{Back-action on the mechanical motion}

Overlaid on the skewed step function are clear minima of the resonance frequency whenever
single electron tunneling takes place. In addition to static and radio frequency driving
force, now the fluctuation of the electric charge contributes to the motion. As presented
in more detail in \cite{strongcoupling}, this dynamical effect adds an effective softening
spring constant. This leads to dips in the resonant frequency up to an order of magnitude
greater than the shifts induced by one more elementary charge on the nanotube. The
resulting ``Coulomb blockade oscillations of mechanical resonance frequency'' are 
illustrated in the measurement of Fig.~\ref{fig-back}(c). Here the mechanical resonance
can be traced across three Coulomb blockade oscillations, with the characteristic
repetition of the pattern of Figs.~\ref{fig-back}(a,b).

To describe the underlying mechanism we need to point out that at finite temperature and
finite dot--leads coupling the charge carrier number on the quantum dot fluctuates by one
over a finite potential region. Across this region, the time-averaged charge
$\overline{q}$ varies smoothly with the potential. Now we again assume that electronic
tunneling takes place at a much faster timescale than mechanical motion. This means that
$\overline{q}(u)$ is well-defined for each point of the mechanical oscillation cycle.
Motion of the of the resonator towards the gate leads to an increase of the gate
capacitance $\cg(u)$, which influences a larger charge $\overline{q}(u)$ on the resonator,
in turn increasing the force which pulls it towards the gate. The fact that the
electrostatic force on the nanotube depends on the displacement means that an effective
modification of the spring constant takes place.

Quantitatively, the nano-electromechanic spring constant contribution is proportional to
$\text{d} \overline{q}/\text{d}(\vg\cg)$. As detailed in \cite{strongcoupling}, at low
bias the charge on the quantum dot changes over a small gate voltage region, leading to a
narrow and deep dip in the resonance frequency $f(\vg)$. In contrast, at high bias the
charge can change gradually over the single electron tunneling region, resulting in a
broad and shallow response of the mechanical resonance frequency. 

\begin{figure}[ht]%
\includegraphics*[width=\linewidth]{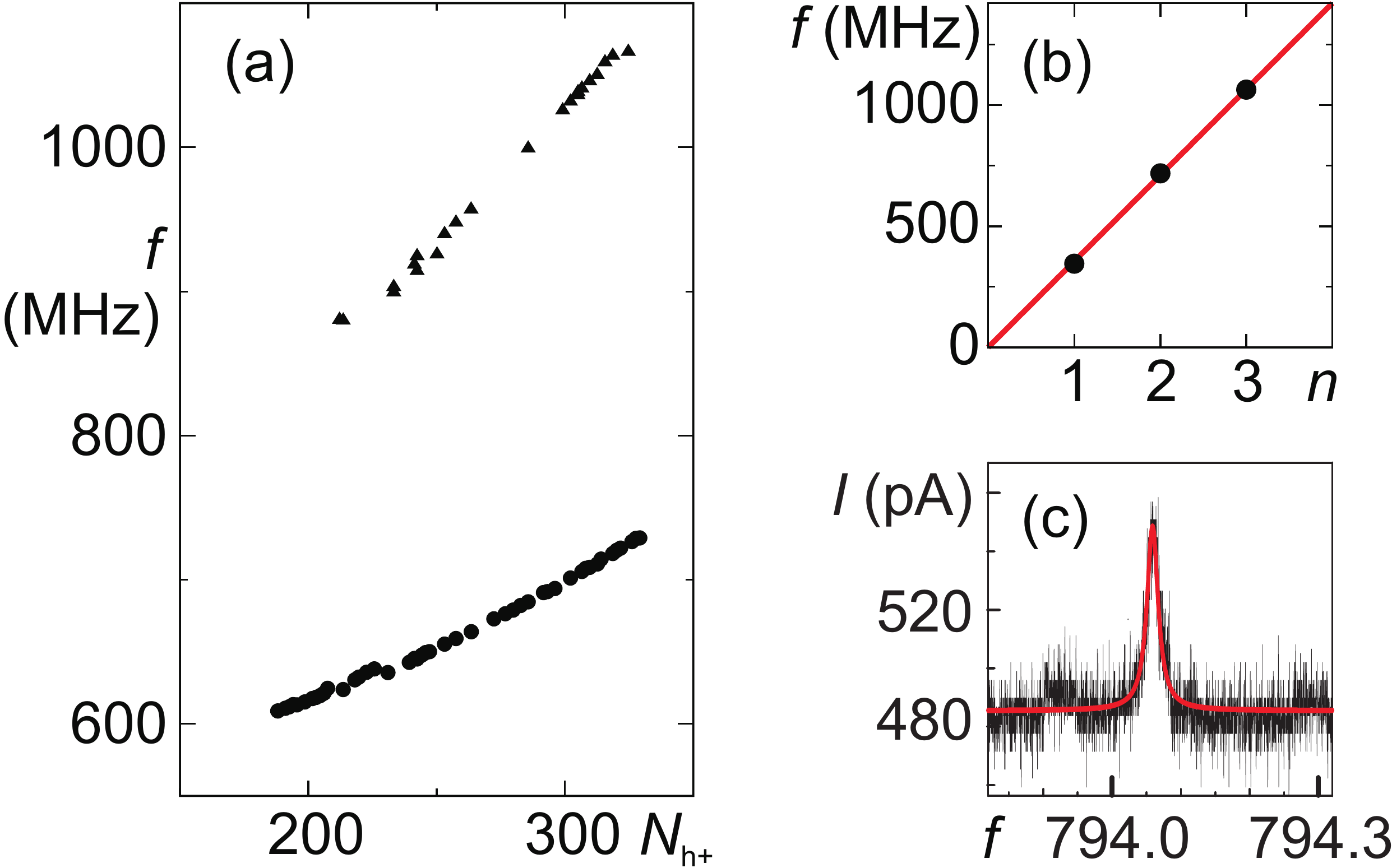}
\caption{%
  (a) Higher frequency mechanical modes: frequency as function of the charge number
  $N_{\text{h}^+}$ on the nanotube, where detected in the dc current signal. 
  (b) Mode frequency plotted as function of the mode number for $N_{\text{h}^+}\simeq
  320$ ($\vg\simeq -6.29\un{V}$), i.e. the high-tension limit. As expected for a string
  under tension the higher mode frequencies ($n=2, 3$) approach integer multiples of
  the fundamental resonance frequency ($n=1$).
  (c) Example trace of the dc current $I_{\text{dc}}(f)$ as function of the driving
  frequency $f$, demonstrating detection of the first higher vibration mode at
  $\vg\simeq-7.5\un{V}$. The solid red line displays the result of a curve fit with
  parameters $f_0'=794.06\un{MHz}$ and $Q=40406$.}
  \label{fig-higher}
\end{figure}
\section{Higher mechanical resonator modes and outlook}

Already in the first measurements on nanotube beam resonators not only the fundamental
flexural mode, but also other mechanical resonances were observed
\cite{nature-sazonova:284}. The simplicity of the radio-frequency part of our setup
enables us to transfer this to significantly higher frequencies. This is illustrated in
Fig.~\ref{fig-higher}(a), where two higher mechanical modes are traced over a large range
of nanotube charge and thereby mechanical tension. The obtained frequency values at
$N_{\text{h}^+}\simeq 320$ ($\vg\simeq -6.29\un{V}$), i.e. in the high-tension limit, are
plotted in Fig.~\ref{fig-higher}(b) as function of the mode number $n$. As expected for a
string under high mechanical tension, the higher vibration mode frequencies ($n=2,3$)
approach integer multiples of the fundamental vibration mode frequency ($n=1$). This also
confirms that we indeed observe the fundamental mode and the second and third
flexural mode of the nanotube resonator.

Fig.~\ref{fig-higher}(c) displays a corresponding current signal $I_{\text{dc}}(f)$ at low
driving power, with a symmetric frequency response as expected for a harmonic oscillator.
A mechanical response can be found at frequencies of $f \simeq 1\un{GHz}$, and even
at higher frequencies signals may still be discovered through careful filtering of the raw
measurement data. 

The frequency scale of $1\un{GHz}$ is significant insofar as it corresponds via $h f
=k_\text{B} T$ to a temperature of $T\simeq 50\un{mK}$, which can easily be obtained with
state-of-the-art dilution refrigerators. This way, the transition between a ``classical''
nanotube beam resonator and a quantum mechanical harmonic oscillator comes into
experimental reach \cite{Schwab2005Putting,nature-oconnell:697}.
Higher resonator modes are however not the only method of approaching
the quantum regime. The bending vibration mode frequency depends strongly on the length
$L$ of the resonator, scaling with $1/L^2$. Current nanofabrication techniques enable
definition of structures significantly smaller than the resonator length $L=800\un{nm}$
of the device presented here. While the greater challenge for
quantum-nano-electromechanical systems will likely lie in the development of driving-free
sensitive motion detection schemes, nanotube resonators provide a highly attractive
material system and may well be at the core of many interesting future experiments.

{\small {\bf Acknowledgments.} This research was carried out with financial support from the Dutch Foundation for
Fundamental Research on Matter (FOM), The Netherlands Organisation for Scientific Research
(NWO), NanoNed, and the Japan Science and Technology Agency International Cooperative
Research Project (JST-ICORP). A.~K.~H. acknowledges financial support by the Deutsche
Forschungsgemeinschaft (DFG) via SFB 689/A1.}

\bibliography{paper}

\end{document}